\documentclass[notitlepage,
 twocolumn,
superscriptaddress,
nofootinbib,
 amsmath,amssymb,
 aps,
 prb]{revtex4-2}
\usepackage{amsmath,amssymb,graphicx,mathtools,dsfont,bbold, natbib}
\usepackage[dvipsnames]{xcolor}
\usepackage[hidelinks]{hyperref}
\usepackage{soul}
\usepackage{bibunits}
\usepackage{dcolumn}% Align table columns on decimal point
\usepackage{bm}% bold math
\usepackage{color}
\usepackage{physics}
\usepackage{dsfont}
\usepackage{adjustbox}
\usepackage{soul}
\usepackage{multirow}
\usepackage{rotating} % Rotating table
\usepackage{comment}

\newcommand{\IOPPAS}{Institute of Physics PAS, Aleja Lotnik\'ow 32/46, 02-668 Warszawa, Poland}
\newcommand{\UW}{Institute  of Theoretical Physics, Faculty of Physics, University of Warsaw, Pasteura 5, 02-093 Warsaw, Poland}

\begin{document}

\title{
Magnetization-induced reordering of ground states phase diagram in a two-component Bose-Hubbard model
}

\author{Oskar Stachowiak$^*$}
\affiliation{\IOPPAS}
\affiliation{\UW}

\author{Hubert Dunikowski$^*$}
\affiliation{\IOPPAS}
%\orcid{0009-0004-0076-645X}
%\email{dunikowski@ifpan.edu.pl}

\author{Emilia Witkowska}
\affiliation{\IOPPAS}
%\orcid{0000-0003-4622-8513}
\email{ewitk@ifpan.edu.pl}

\date{\today}

\begin{abstract}
We investigate the influence of non-zero magnetization on the ground-state phase diagram of the two-component Bose-Hubbard model.
Employing a mean-field theoretical framework, both analytically and numerically, we demonstrate that positions and sizes of specific phases on the diagram are magnetization dependent.
In particular, non-zero magnetization introduces different Mott insulator phase boundaries for each of the two components.
This effect leads to the emergence of a hybrid phase characterized by the coexistence of superfluid in one of the components and Mott insulator in the another one.
Our findings highlight the important role of a conserved quantities, which is magnetization here, in reshaping the phase landscape, significantly influencing the stability and emergence of distinct quantum phases.
\end{abstract}

\maketitle

\def\thefootnote{*}\footnotetext{These authors contributed equally to this work}\def\thefootnote{\arabic{footnote}}

\section{Introduction}
\label{sec:Introduction}

The Hubbard model, introduced in 1963 to describe electrons in solids~\cite{PhysRev.129.959, Hubbard1963, Hubbard1964, Hubbard1965}, explains the hopping of particles between the lowest vibrational states of lattice sites~\cite{Montorsi1992, Pavarini2017}. 
This mathematical framework has led to a deep understanding of various magnetic phases, including ferromagnetic and antiferromagnetic~\cite{Majlis2007, Pavarini2015} phases, as well as exotic phases like spin liquids~\cite{Savary2017}. Consequently, it has provided valuable insights into the realm of quantum magnetism~\cite{schollwock2004quantum}.

Over the last twenty-five years, there has been intense interest in studying Hubbard physics with ultra-cold atoms in optical lattices, both experimentally and theoretically~\cite{Bloch2012}. 
In addition to the standard superfluid (SF) and Mott insulator (MI) phases at low temperatures, there are other even more exotic phases for bosonic and fermionic atoms with extended models involving long-range interactions, next-nearest-neighbor or frustrated hopping terms~\cite{Lewenstein2012, Dutta_2015, Kovrizhin_2005}. For example, charge density waves, high-temperature superconductivity, topological phases, ferromagnetic and spin liquid phases~\cite{Aidelsburger2013, PhysRevLett.108.255303}, and striped phases~\cite{Bourgund2025} are appearing, some of them were observed experimentally as well.
Nowadays, new opportunities are emerging in the study of exotic phases of Hubbard models using systems of Rydberg atom arrays ~\cite{weckesser2024realizationrydbergdressedextendedbose, Browaeys2020, Moses2017}, trapped ions~\cite{Britton2012} and molecules~\cite{Gorshkov2011, Tarruell2018}.

While the single-component Hubbard model has been extensively studied, the natural extension to a binary mixture of bosons with two internal states $\ket{a}$ and $\ket{b}$ (where these two species define a pseudo-spin) in an optical lattice presents new opportunities for investigation.
In particular, the two-component Bose-Hubbard (TBH) model, characterised by nearest-neighbour tunnelling $J$, intra-species $U$ and inter-species $U_{ab}$ on-site interactions ~\cite{Altman_2003, PhysRevLett.92.050402}, exhibits a rich phase diagram ~\cite{Altman_2003, PhysRevLett.92.050402}.
Notably, it includes exotic phases of strongly correlated states such as a counterflow superfluid (CFSF) and a pair superfluid (PSF)~\cite{PhysRevLett.90.100401, PhysRevA.80.023619, PhysRevA.80.023619, 10.21468/SciPostPhys.12.3.111, PhysRevLett.128.093401}. These phases are superpositions of MI states of different occupations of the spin levels, such that they reveal superfluidity in the internal spin degree of freedom. In CFSF the total occupation of a site is fixed, so that the numbers of atoms in components in a site are anticorrelated.
Inversely, in PSF the occupation of a site differs, between MI states forming the superposition, by a pair of atoms -- one in each component, what gives rise to correlation in atom numbers in components in a site.

We investigate how fixed magnetization, which is an on-site population imbalance, and the interspecies interaction strength $U_{ab}$ shape the phase diagram of ground states. Employing mean-field framework we performed analytical perturbative calculations to determine boundaries between various phases and numerical self-consistent procedure to map the phase diagram of the ground states.

We derived simple analytical expressions for the SF–MI and CFSF–MI phase boundaries where interactions are purely repulsive $U\ge U_{ab}\ge 0$.
The CFSF phases emerge as lobs nested between successive MI lobes, apart from the situations when $U_{ab}=0$ or $U_{ab}=U$ where there are MI or CFSF lobes only, respectively.
We show that introducing an integer odd magnetization value changes the character of the lobs from MI to CFSF, and vice-versa. However, in the low chemical potential, the CFSF phase disappears.
In addition, lobs sizes can be different for $a$ and $b$ components for the same chemical potential and non-zero magnetization.  
We identify a parameter range where the component $a$ is in the SF phase while the component $b$ in the MI phase, and vise versa, leading to a coexistence of SF and MI behavior within the one system.
In the case of attractive interspecies interactions $(U_{ab}<0)$, we derive the critical chemical potential at which the pair-superfluid (PSF) phase emerges as a boundary between neighboring MI lobs.
Although our mean-field treatment may not capture every subtlety of considered transitions, it nevertheless outlines the key features and critical points governing the system’s overall phase structure.

Our findings highlight the crucial role of conserved quantities in determining the phase structure of ground states of a lattice system. The presence of such a conserved quantity can substantially modify the phase diagram, affecting both the stability and existence of individual quantum phases. Specifically, in the system under investigation, the CFSF phase vanishes at low chemical potentials when the magnetization is non-zero.
At sufficiently large chemical potential, the interplay between the conserved quantity and interaction parameters leads to a reordering of the phase landscape.

\section{The model}
\label{sec:The model and methods}

We consider $N$ bosonic ultra-cold atoms with two internal states $\ket{a} $ and $\ket{b}$ loaded in the optical lattice potential, described by the two-component $d$-dimensional Bose-Hubbard Hamiltonian,
\begin{align}
        \hat{\mathcal{H}} = &
        - J  \sum\limits_{\langle i,j \rangle}
        (\hat{a}_{i}^{\dagger}\hat{a}_{j} + \hat{b}_{i}^{\dagger}\hat{b}_{j})
        + U_{ab} \sum_i 
        \hat{n}_{a,i} \hat{n}_{b,i}
        \nonumber \\
        + &  \frac{U}{2}
        \sum\limits_{i, \alpha=a,b}
        \hat{n}_{\alpha,i}(\hat{n}_{\alpha,i} - 1)
        - \sum\limits_{i, \alpha=a,b} \mu_{\alpha} \hat{n}_{\alpha,i},
        \label{eq:Ham0}
\end{align}
with periodic boundary conditions.
$\hat{a}_{i}$ and $\hat{b}_{i}$ are on-site annihilation operators of boson in the state $a$ and $b$, respectively. 
The local number operators are $\hat{n}_{a,i} = \hat{a}_i^{\dagger} \hat{a}_i$ and $\hat{n}_{b,i} = \hat{b}_i^{\dagger} \hat{b}_i$. Additionally, $\mu_{a}$ and $\mu_{b}$ are the chemical potentials for the $a$ and $b$ components, imposing their occupations independently. 
The tunnelling $J$, inter- and intra-species interaction energies $U$ and $U_{ab}$ depend on details of the optical lattice potential~\cite{Jaksch1998-yn}. 
The above Hamiltonian, conserves the number of bosons in the $a$ and $b$ components independently. Therefore, the magnetization, which is defined as a difference in the population number $M=N_a - N_b$, is a constant of motion. This motivates studying the ground states phase diagram corresponding to a given magnetization sector. 

We focus exclusively on uniform systems, considering the interplay between interactions and magnetization, and their impact on the boundary between various phases. 
Each phase constituting the ground state phase diagram can be characterized by an appropriate order parameter. 
In the case of the SF and MI phases the convenient order parameters are $\varphi_{a,i} = \langle \hat a_{i}^\dagger \rangle$ for the component $a$, and $\varphi_{b,i} = \langle \hat b_{i}^\dagger \rangle$ for the component $b$. 
These parameters are non-zero in the respective SF phase only. 
A change of their values to zero, $\varphi_{a,i}=\varphi_{b,i} =0$, indicates a transition from the SF to the MI, CFSF or PSF phase.
In addition, we introduce CFSF and PSF order parameters,
\begin{align}
    \varphi_{c, i} =& \langle\hat a_{i}\hat b^{\dagger}_{i} \rangle - \langle\hat a_{i} \rangle \langle\hat b^{\dagger}_{i} \rangle, \\
    \varphi_{p, i} =& \langle\hat a_{i}\hat b_{i} \rangle - \langle\hat a_{i} \rangle \langle\hat b_{i} \rangle,
\end{align}
respectively, which are nonzero in the CFSF and PSF phases, differentiating them from the MI.

\section{Analytical results: mean-field perturbative approach}
\label{subsec:Analytical perturbative approach}

We employ a perturbative mean-field (MF) framework to map the phase diagram of the ground states on the $(J,\mu)$ plane.
In this method, we disregard the second-order fluctuations of the bosonic annihilation and creation operators, assuming $(\hat{a}^{\dagger}_{i} - \langle\hat{a}^{\dagger}_{i}\rangle)(\hat{a}_{j} - \langle\hat{a}_{j}\rangle) \approx 0$ for $a$ and $(\hat{b}^{\dagger}_{i} - \langle\hat{b}^{\dagger}_{i}\rangle)(\hat{b}_{j} - \langle\hat{b}_{j}\rangle) \approx 0$ for $b$. 
In the homogeneous case, we can assume the superfluid parameters $\varphi_{a} = \langle\hat{a}^{\dagger}_{i}\rangle$ and $\varphi_{b} = \langle\hat{b}^{\dagger}_{i}\rangle$ to be lattice site independent and real, without loss of generality.
The Hamiltonian (\ref{eq:Ham0}) under such assumptions is a simple sum, 
\begin{equation}
\hat{\mathcal{H}}^{\text{MF}} = \sum_{i}\hat{h}_{i},
\end{equation}
of local Hamiltonian terms
\begin{align}
\hat{h}_{i} 
&
= 
\sum_{\alpha \in \{ a,b \}} \Big[ \frac{U}{2}\hat{n}_{\alpha,i}(\hat{n}_{\alpha i}-1) - \mu_{\alpha} \hat{n}_{\alpha,i} \nonumber \\
&+ zJ\varphi_{\alpha}^2 - zJ\varphi_{\alpha}( \hat{\alpha}^{\dagger}_{i} + \hat{\alpha}_{i} )\Big] +U_{ab}\hat{n}_{ia}\hat{n}_{ib},
\label{eq:localhamiltonian}
\end{align}
where $z$ denotes the number of nearest neighbors. 
In what follows, we simplify our analysis by focusing on the local Hamiltonian (\ref{eq:localhamiltonian}) only; therefore, we drop the indices corresponding to the lattice sites. 
 
To perform the perturbative analysis of the phase boundaries, the local Hamiltonian (\ref{eq:localhamiltonian}) is considered as a sum $\hat{h} = \hat{h}_{0} + \hat{v}(J)$ of the unperturbed term
\begin{equation}
\hat{ h }_{0} = 
\sum_{\alpha \in \{ a,b \}} 
\left[
\frac{U}{2}\hat{n}_{\alpha}(\hat{n}_{\alpha}-1) - \mu_{\alpha} \hat{n}_{\alpha} + zJ\varphi_{\alpha}^2
\right] +U_{ab}\hat{n}_{a}\hat{n}_{b},
\label{eq:unperturbH}
\end{equation}
and the perturbation
\begin{equation}
\hat{v}(J) = - zJ\varphi_{a}( \hat{a}^{\dagger} + \hat{a} ) - zJ\varphi_{b}( \hat{b}^{\dagger} + \hat{b} ).
\label{eq:perturbation}
\end{equation}
The unperturbed local energy reads
\begin{equation}
    \begin{aligned}
        E_{0}^{(0)}(n_a,n_b) =& \sum_{\alpha \in \{ a,b \}} \left[ \frac{U}{2}n_{\alpha}(n_{\alpha}-1) - \mu_{\alpha}n_{\alpha} + zJ \varphi_{\alpha}^2 \right] \\
        &+U_{ab}n_{a}n_{b},
    \end{aligned}
    \label{eq:GSEnergy}
\end{equation}
when considering (\ref{eq:unperturbH}) in the local Fock eigenbasis \( |n_{a},n_{b}\rangle \). Here, \( n_a \) and \( n_b \) are integers representing the number of atoms in \( a \) and \( b \), respectively.

Before proceeding with further analysis, we recast the zeroth‑order onsite energy \( E_{0}^{(0)} \) in terms of the local number of atoms \(n = n_a + n_b\) and the local magnetization \(m = n_a - n_b\). Additionally, we introduce the average chemical potential $\mu =(\mu_a + \mu_b)/2$ and the chemical potential difference $h=(\mu_a - \mu_b)/2$ (which acts as an effective magnetic field). With these variables, 
\begin{align}
E_{0}^{(0)}(n,m) = &
n^2 \frac{U+U_{ab}}{4} + m^2 \frac{U-U_{ab}}{4} \nonumber \\
-&n \left(\mu +\frac{U}{2}\right)-m h + z J (\varphi_{a}^2 + \varphi_{b}^2)\;.
\end{align}
Note that in this equation, \(n\) and \(m\) are both integers. The local energy $E_{0}^{(0)}(n,m)$ can be seen as a quadratic function of the local magnetization $m$, for the fixed local number of atoms $n$, with a minimum at $m_{min} = 2h/D$, where $D=U-U_{ab}$. 
Although we treat $m_{min}$ as a continuous variable in this context, it can only take discrete values in the system, as it is related to a specific value of magnetization. 
Consequently, we infer that the ground state of equation (\ref{eq:unperturbH}) becomes degenerate for various \( n \) only when \( h \) is a multiple of \( \frac{D}{2} \). Thus, we introduce an integer \( l \) such that $h = l D/2$, which is the value of magnetization at energy minima Therefore, we consider the ground states parametrized by $l$.

The interplay between parities of $n$ and $l$ plays a crucial role in determining the ground state. Specifically, the MI phase occurs for an even total number of particles $n$ when $l$ is even. Conversely, when $l$ is odd, the MI phase appears for an odd number of particles.
The ground state in both cases reads
\begin{equation}
|\Psi_{\rm MI}\rangle = 
\ket{ \frac{n + l}{2}, \frac{n - l}{2}},
\end{equation}
where $n$ and $l$ share the same parity.
On the contrary, when $n$ and $l$ do not have the same parity, the ground state becomes degenerate with the two states:
\begin{align}
\label{eq:X1}
\ket{X_1}&=\ket{\frac{n + l-1}{2}, \frac{n - l +1}{2} },\\
\label{eq:X2}
\ket{X_2}&=\ket{\frac{n + l +1}{2}, \frac{n - l-1}{2} },
\end{align}
indicating the CFSF phase's presence~\cite{PhysRevLett.90.100401, 10.21468/SciPostPhys.12.3.111}. In this case, we consider the ground state as the symmetric superposition
\begin{equation}
|\Psi_{\rm CF}\rangle
= 
\frac{1}{\sqrt{2}} \Big(
\ket{X_1}
+ 
\ket{X_2}
\Big).\label{eq:Psi_CF}
\end{equation}

\subsection{Transition from MI to SF}
\label{subsec:MItoSF}

We start the analysis with the focus on the boundaries between the MI and SF phases.
Since the perturbation given by Eq.~(\ref{eq:perturbation}) includes terms that are linear in the creation and annihilation operators, specifically 
$\hat{a}^{\dagger} + \hat{a}$ and $\hat{b}^{\dagger} + \hat{b} $, only even terms in the perturbation series will contribute. 
For the non-degenerate case, 
the ground-state energy up to second order in perturbation is given by
\begin{equation}
E_{0} \simeq 
E_{0}^{(0)} + E_{0}^{(2)},
\end{equation}
with
\begin{equation}
    E_{0}^{(2)} =  
    \sum_{n_a, n_b}
    \frac{| \langle n_a, n_b | \hat{v} | \Psi_{\rm MI} \rangle |^2}{E_{0}^{(0)} - E_0^{(0)}(n_a, n_b)}.
\end{equation}
This gives
\begin{align}
E_{0} &= 
\sum_{\alpha \in \{ a,b\}}\left[ \frac{U_{\alpha}}{2} n_{\alpha} (n_{\alpha}-1) - \mu_{\alpha}n_{\alpha} + z J r_{\alpha} \varphi_{\alpha}^{2} \right]  \nonumber \\
&+ U_{ab} n_{a} n_{b},
\end{align}
with
\begin{align}
    r_{a} &=  1 - zJ
    \frac{ 2(n U_{ab} -  2\mu ) - 2(l+2) U  }
    {\left(n U_+ - 2U_\mu\right)\left(nU_+- 2\mu \right) },\\
    r_{b} &=  1 - zJ
    \frac{  2(n U_{ab} - 2 \mu ) + 2(l-2) U  }
    {\left(nU_+ - 2U_\mu \right)(nU_+- 2\mu ) },
\end{align}
when introducing $U_{+}=U + U_{ab}$, $U_\mu=U+\mu$.

If $r_{\alpha} > 0$, the corresponding order parameter 
$\varphi_{\alpha} = 0$ minimizes the energy, indicating the presence of the MI phase in the $\alpha$ component, denoted as ${\alpha}$-MI. Conversely, when $r_{\alpha} < 0$, the energy is minimized for $\varphi_{\alpha} \neq 0$, which indicates a superfluid phase in the component $\alpha$. 
Therefore, the boundary separating the MI and SF phases for the $\alpha$ component is defined by the condition $r_{\alpha} = 0$.
The corresponding analytical formulas for the $(J,\mu)$ plane read
\begin{equation}
    z J^{(a-{\rm MI})}_c=
    \frac{\left(n U_+ - 2U_\mu\right)\left(nU_+- 2\mu \right) }
    {  2(n U_{ab} -  2\mu ) - 2(l+2) U  },
    \label{eq:MIa}
\end{equation}
for the component $a$, and 
\begin{equation}
    z J^{(b-{\rm MI})}_c=
   \frac{\left(nU_+ - 2U_\mu \right)(nU_+- 2\mu ) }
   {  2(n U_{ab} - 2 \mu ) + 2(l-2) U  },
   \label{eq:MIb}
\end{equation}
for the component $b$.

\subsection{Transition from CFSF to SF}

As we transition to the discussion of the CFSF phase, it is important to emphasize that this phase can only occur in the system for finely tuned values of \( h \), namely \( h = \frac{l D}{2} \), where \( l \in \mathbb{Z} \). Within CFSF phase, the local number of atoms \( n \) has a different parity than \( l \). Moreover, the states $\ket{X_1}$ and $\ket{X_2}$, as defined in Eqs.~(\ref{eq:X1}) and (\ref{eq:X2}), respectively, are eigenstates of unperturbed Hamiltonian. 
For a fixed local magnetization, the ground state $| \Psi_{\rm CF} \rangle$ is the symmetric superposition of $\ket{X_1}$ and $\ket{X_2}$.  In the degenerate perturbation theory that follows, it is therefore convenient to introduce the orthogonal “antisymmetric” combination,
\begin{equation}
    | \Psi_\perp \rangle=\frac{1}{\sqrt{2}} \Big( \ket{X_1} - \ket{X_2} \Big),
\end{equation}
 which, together with $| \Psi_{\rm CF} \rangle$, spans the two-dimensional subspace of interest. 
To determine the energy shifts to second order in the perturbation, we construct the matrix of second-order corrections,
\begin{align}
    \sum_{k} 
    \left[ \begin{array}{cc}
     \frac{\langle \Psi_{\rm CF} | \hat{v}(J)| k \rangle
     \langle k| \hat{v}(J) | \Psi_{\rm CF} \rangle}{E_{0}^{(0)} - E^{(0)}(k)} & \frac{\langle \Psi_{\rm CF} | \hat{v}(J) | k \rangle
     \langle k| \hat{v}(J) | \Psi_\perp \rangle}{E_{0}^{(0)} - E^{(0)}(k)}  \\ \frac{\langle \Psi_\perp | \hat{v}(J) | k \rangle
     \langle k| \hat{v}(J) | \Psi_{\rm CF} \rangle}{E_{0}^{(0)} - E^{(0)}(k)} & \frac{\langle \Psi_\perp | \hat{v}(J) | k \rangle
     \langle k| \hat{v}(J) | \Psi_\perp\rangle}{E_{0}^{(0)} - E^{(0)}(k)} \end{array} \right],
     \label{eq:matrix}
\end{align}
where $k=\{n_a, n_b\}$. 
Given that $| \Psi_{\rm CF} \rangle$ denotes the ground state of the system and $| \Psi_{\perp} \rangle $ is orthonormal to it, the matrix defined in Eq. (\ref{eq:matrix}) takes a diagonal form. 
This implies that the off-diagonal elements of the matrix vanish,
\begin{equation}
\sum_{k} \frac{\langle \Psi_{\rm CF} | \hat{v}(J) | k \rangle
     \langle k| \hat{v}(J) | \Psi_{\perp} \rangle}{E_{0}^{(0)} - E^{(0)}(k)} =0,
\end{equation}
establishing a relationship between the order parameters $\varphi_{a}$ and $\varphi_{b}$ of the form 
\begin{equation}
 \varphi_{b}^{2} 
 = \frac{n+1+l}{n+1-l}\,\varphi_{a}^{2}   \label{eq:phi_a_phi_b_CFSF_ratio}.
\end{equation}
Applying this relation to the diagonal elements yields the system's energy up to second order in perturbation theory, given by:
\begin{equation}
E_{0} = \sum_{\alpha \in \{ a,b \}}\left[\frac{U_{\alpha}}{2}n_{\alpha}(n_{\alpha}-1)\right]
+U_{ab}n_{a}n_{b} + zJr_{\rm CF}\varphi_{a}^{2},
\end{equation}
with
\begin{equation}
    r_{\rm CF} = 1 - zJ \frac{1}{2(n+1)} \frac{A}{B},\label{eq:CFSF}
\end{equation}
where $B = (x_\mu^2 - D^2 ) (x_\mu-U_+)(x_\mu -U_{+} - 2 D)$ and
\begin{align}
    & A  = 
    4 \left[   l^2 - (n+1)^2 \right] D^3 + \left[l^2 - (n+ 1) (n+7)\right] x_\mu^2 D \nonumber \\
    &+ \left[l^2 - (n+1) (n+3)\right] (U_+ - x_\mu ) D^2  \nonumber \\
    &+ \left[\left( 3 l^2 - (n{+} 1 ) (3 n{+} 5) \right) U_+\! + 4 (n{+}1) x_\mu \right] x_\mu (x_\mu - U_+)
   \nonumber \\
    &+(n^2 -l^2 - 1)  U_+^2 D  +4 (n^2  - l^2 + 4 n + 3) x_\mu U_+  D,\nonumber
\end{align}
and $x_\mu =n (U+U_{ab}) - 2\mu$.

Analogous to the non-degenerate case discussed in Sec.~\ref{subsec:MItoSF}, when $r_{\rm CF} > 0$, the energy is minimized by setting the order parameter to zero, $\varphi_{a} = 0$, indicating that the system resides in the CFSF phase. Conversely, for $r_{\rm CF} < 0$, the energy is minimized for a non-zero order parameter, $\varphi_a \neq 0$, marking the emergence of the SF phase. Consequently, the phase boundary between the CFSF and SF phases is determined by the condition $r_{\rm CF} = 0$.
The explicit analytical expression for the CFSF–SF phase boundary in the $(J, \mu)$ plane is:
\begin{equation}
zJ_c^{(\rm CF)} = 2(n+1) \frac{B}{A}.
\label{eq:JcCFSF}
\end{equation}

\section{Numerical results and discussion}
\label{sec:Numerical self-consistent approach}
%%%%%%%%%%%%%%%%%%%%%%%%%%

\begin{figure*}[hbt!]
\includegraphics[width=\linewidth]{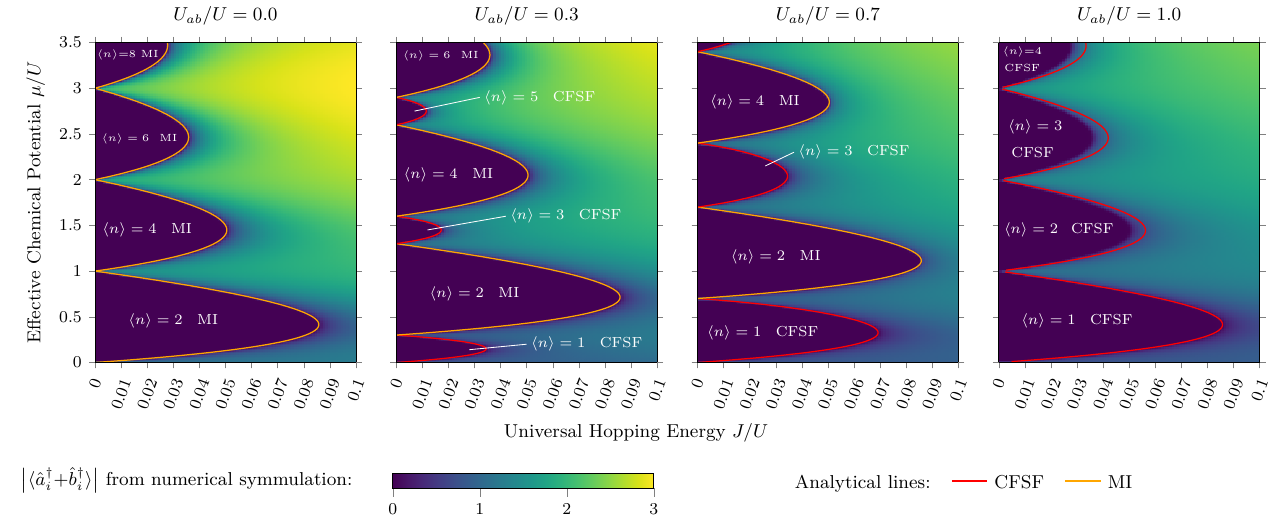}
\caption{
Ground-state phase diagrams are presented for varying interspecies interaction strength $U_{ab}/U=0,\, 0.3,\, 0.7,\, 1$ (from left to right), in the absence of an effective external field ($h=0$) corresponding to the zero magnetization value. The color maps depict the numerically computed superfluid order parameter \( \varphi_a + \varphi_b \). Solid lines correspond to analytically derived phase boundaries, given by Eqs.~(\ref{eq:MIa}), (\ref{eq:MIb}), and (\ref{eq:JcCFSF}), delineating the transitions between distinct quantum phases.
}
\label{fig:fig1}
\end{figure*}

%%%%%%%%%%%%%%%%%%%%%%%%%%%%

We employ a mean-field self-consistent (MFSC) numerical procedure to investigate the ground-state phase diagram of the system. In this approach, we consider the local Hamiltonian $\hat{h}_i$ describing a single site, as defined in Eq.~\eqref{eq:localhamiltonian}. The procedure begins with an initial trial state from which the order parameters $\varphi_a$ and $\varphi_b$ are computed. Using these values, we construct and diagonalize the local Hamiltonian $\hat{h}_i$ to obtain an updated ground-state wavefunction. The corresponding updated values of $\varphi_a$ and $\varphi_b$ are then reinserted into $\hat{h}_i$. These steps of procedure are iterated until self-consistency is achieved.

\subsubsection{Effect of $U_{ab}$ (for $h=0$)}

Fig.~\ref{fig:fig1} illustrates the ground-state phase diagrams for equal chemical potentials, $h = 0$, emphasizing the effect of varying interspecies interaction strength $U_{ab}$. 
Here, the numerical MFSC method converges to a solution with zero local magnetization $m=0$ in the whole range of $J$ and $\mu$.
A strong agreement between the analytically and numerically determined phase boundaries of the MI and CFSF phases is observed. 

In the absence of interspecies coupling ($U_{ab} = 0$), the system consists of two decoupled, non-interacting bosonic species, $a$ and $b$, each exhibiting identical phase diagrams. The MI lobes of the two components overlap, yielding insulating regions with even total occupations only.
As $U_{ab}$ increases, CFSF lobes characterized by odd average total occupations emerge between the even-occupation MI lobes. This behavior reflects the formation of interspecies pairing correlations enabled by the repulsive interaction.
With further increase of $U_{ab}$, the CFSF regions broaden, indicating the enhanced stability of the CFSF phase due to stronger interspecies coupling.
In the limit $U_{ab} = U$, the interaction Hamiltonian looses distinction between components $a$ and $b$, as all interaction strengths become equal. Under these conditions only the total density $n = n_a + n_b$ remains well-defined inside the lobs (where where tunneling is suppressed), allowing arbitrary mixtures of the two components within each lobe. Because of that in this regime all insulating lobes can accommodate the CFSF phase for $m=0$, as observed in the phase diagrams.

The emergence and growth of CFSF lobes with increase of $U_{ab}$ reflect enhanced intercomponent correlations. This behavior is directly relevant for ultracold atoms experiments, where interaction tunability enables controlled access to such correlated quantum phases~\cite{PhysRevLett.128.093401}.

%%%%%%%%%%%%%%%%%%%%%%%%%%%%%%%%%%%%

\subsubsection{Effect of $h=lD/2$ for $l\in\mathbb{N}_+$}
\label{sec:fixed_h}

\begin{figure*}[]
\includegraphics[scale=1.0]{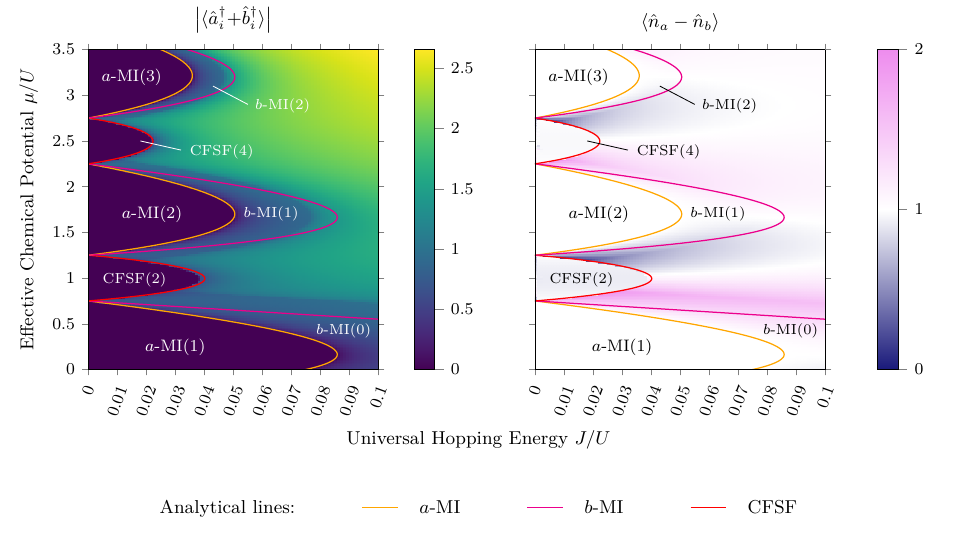}
\caption{
Left panel: The ground state phase diagram for nonzero magnetization when $h=D/2$ and $U_{ab}/U=0.5$. 
Color map indicate the order parameter values $\varphi_a + \varphi_b$ from numerical simulation and are combined with analytical results for the loop boundaries, Eqs. (\ref{eq:MIa}), (\ref{eq:MIa}) and (\ref{eq:JcCFSF}). 
Right panel: Ground-state magnetization values, $m = n_a - n_b$, obtained from numerical simulations. Results demonstrate that, unlike in the $a$-MI and CFSF phases, where $m=1$, the numerical self-consistent procedure does not keep this magnetization value when at least one component is in the SF phase.
}
\label{fig:fig2}
\end{figure*}
%%%%%%%%%%%%%%%%%%%%%%%%%%%%%%%%%%%%%%%%%%%%%

%%%%%%%%%%%%%%%%%%%%%%%%%%%%%%%%%%%%%%%%%%%%%
\begin{figure*}[]
\includegraphics[scale=1.0]{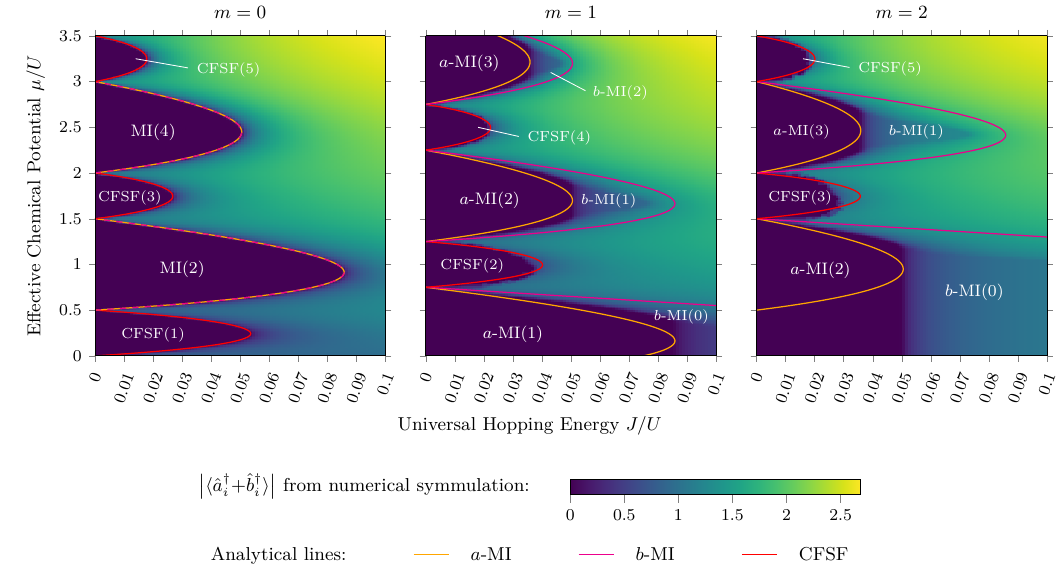}
\caption{
Ground-state phase diagrams for fixed values of on-site magnetization to $m = 0,\, 1,\, 2$ (from left to right), obtained from numerical self-consistent simulations incorporating a bisection procedure. Analytical phase boundaries, derived from Eqs.~\eqref{eq:MIa}, \eqref{eq:MIb}, and \eqref{eq:JcCFSF}, are marked by solid lines.
}
\label{fig:fig3}
\end{figure*}

%%%%%%%%%%%%%%%%%%%%%%%%%%%%%%%%%%%%%%%%%%%%%

Fig.~\ref{fig:fig2} presents results for the first nontrivial value of the effective external field, $h = D/2$, in the Hamiltonian \eqref{eq:localhamiltonian} where degeneracy of Fock states \eqref{eq:X1} and \eqref{eq:X2} can occur, and CFSF state \eqref{eq:Psi_CF} can emerge. 
In contrast to the $h = 0$ case shown in Fig.~\ref{fig:fig1}, where the CFSF phase emerges at odd occupation numbers, these phase here occurs at even occupation numbers. 
Moreover, both numerical simulations and analytical calculations reveal distinct MI–SF transition points for the $a$ and $b$ components, indicating species-dependent critical behavior induced by magnetization.

Analytical and numerical results for the $a$-MI and CFSF lobes show excellent agreement. However, for the $b$-MI lobes, the two approaches yield visibly different outcomes. This discrepancy arises because the analytical results are derived from perturbation theory around the double MI state, where both components $a$ and $b$ reside in their respective MI phases. 
Since $a$-MI lobes directly border the double MI region their shape predicted by perturbative expression is highly exact.
In contrast, the $b$-MI lobes are located in regions where the $a$ component has already undergone the SF transition -- far from the state around which perturbation is performed, so that the second order expansion is not sufficient.
Consequently, the analytical method produces inaccurate predictions for the $b$-MI lobes, while the numerical results remain reliable. Nonetheless, despite this limitation, the analytical approach qualitatively captures the presence of the $b$-MI lobes, roughly reproducing their overall shape.

The right panel of Fig.~\ref{fig:fig2} shows the on-site magnetization of the ground state obtained via numerical simulation. It is evident that the magnetization is not uniform across the whole phase diagram but instead varies over a finite range within the SF phase. The numerical method seeks the global ground state irrespective of magnetization constraints, resulting in a magnetization that is fixed only within the $a$-MI and composite CFSF phases.\footnote{By construction the numerical MFSC method do not keep fixed magnetization in the CFSF lobs, converging rather to some arbitrary superposition of \eqref{eq:X1} and \eqref{eq:X2} degenerated states. In order to impose convergence to symmetric superposition \eqref{eq:Psi_CF} characteristic to CFSF phase, without introducing changes to the Hamiltonian, the additional condition of fixing the ratio of $\varphi_a$ and $\varphi_b$ parameters \eqref{eq:phi_a_phi_b_CFSF_ratio} is used in self-consistent procedure when CFSF lobs are calculated. Although in theory in CFSF phase parameters $\varphi_a$ and $\varphi_b$ are zero, in numerical calculations the residual values of $\varphi_a$ and $\varphi_b$ are sufficient to fix the magnetization inside CFSF lobs -- this is the most successful method of imposing magnetization in CFSF found.} The analytical calculations are perturbative expansions around MI or CFSF states with fixed magnetization; however, they do not guarantee magnetization conservation once perturbations are introduced, as the system follows the global energy minimum. All in all, one has to be aware that both methods reproduce shapes of lobes for global ground state regardless of the magnetization in SF phase. 
%

%%%

\subsubsection{Fixed magnetization}

Physically magnetization is a constant of motion of the Hamiltonian \eqref{eq:Ham0}; thus, an experimentally prepared system is expected to realize the ground state constrained within a fixed magnetization manifold, rather than the global energy minimum that may correspond to a different magnetization value. Fig.~\ref{fig:fig3} illustrates the case where the numerical calculations enforce conservation of on-site magnetization across the whole phase diagram by employing a bisection method -- tuning the parameter $h$ until the resulting state attains the desired magnetization value $m$. For comparison, figure also displays results from the analytical approach, which however, as it was mentioned before in Sec.\ \ref{sec:fixed_h}, do not guarantee magnetization conservation in SF phase.
Although the numerical and analytical phase boundaries exhibit noticeable differences in shape, the analytical results generally reproduce the qualitative behavior of the magnetization-constrained system.

\section{Pair Superfluid for $U_{ab}<0$}
\label{sec:}

The MF method we developed can be also applied in the regime of attractive interspecies interactions, \mbox{$U_{ab} < 0$}. In this case, the CFSF phase is absent from the phase diagram, and the PSF phase emerges as the correlated phase.
PSF phase occurs only for a single specific values of chemical potential as a boundary between two neighboring MI lobs with a site total occupation number differing by two.

Examining the formula for the boundary of the MI phases Eqs.~(\ref{eq:MIa}) and~(\ref{eq:MIb}), we see that it intersects the $\mu$ axis at two points: 
$\mu_{-}(n) = \left[n(U + U_{ab}) -U \right]/2$
and 
$\mu_{+}(n) = n(U + U_{ab})/2 $.
By comparing $\mu_{-}$ for $n$ with $\mu_{+}$ for $n+2$ particles, $\mu_{-}(n) = \mu_{+}(n+2)$, we conclude that the MI phases for different numbers of particles intersect for the non-zero value of $J$ only when $U_{ab} < 0$. 
To determine the critical chemical potential $\mu_c$ at which the PSF phase emerges, we identify the point at which the energies of the adjacent Mott-insulating states become degenerate:
$
E_{0}^{(0)} \left( \frac{n+l}{2}, \frac{n-l}{2} \right) = E_{0}^{(0)} \left( \frac{n+2+l}{2}, \frac{n+2-l}{2} \right)$,
which leads to: 
\begin{equation}
    \mu_{c} = \frac{1}{2}\left[n(U + U_{ab})+ U_{ab}\right].\label{eq:PSF}
\end{equation}
Notably, the critical value of the chemical potential, $\mu_c$, is independent of the local magnetization $m$.

\begin{figure}[]
\includegraphics[scale=1.0]{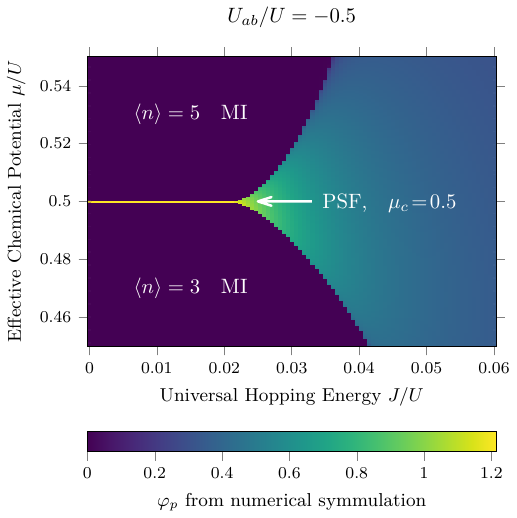}
\caption{
Color map shows the PSF order parameter $\varphi_p$ from numerical simulation results and they are compared with the analytically predicted critical chemical potential, $\mu_c$, at which the PSF phase occurs (indicated by the arrow, calculated from Eq.~\eqref{eq:PSF} with parameters $U_{ab}/U = -0.5$, and total occupation number $n=3$).
}
\label{fig:fig4}
\end{figure}

The emergence of the PSF phase is illustrated in Fig.~\ref{fig:fig4}.
The value of $\mu_c$ predicted by analytical theory is in excellent agreement with the numerical results, which exhibit a pronounced maximum of the paired superfluid order parameter $\varphi_p$ precisely at the expected value of $\mu_c$.

\section{Conclusions and summary}
\label{sec:Conclusions and summary}

We employed the mean-field approach to analyze the ground-state phase diagram of the two-component Bose-Hubbard model. This method reduces the complexity of the many-body problem by decoupling the inter-site tunneling terms, enabling the Hamiltonian to be expressed as a sum of local (on-site) Hamiltonians supplemented by mean-field contributions.

Within this framework, we analytically derived the phase boundaries between the MI, CFSF and SF phases, for arbitrary population imbalance determined by the local magnetization. 
We complemented the analysis with the independent numerical simulations, confirming overall the analytical findings. 
Our results demonstrate how the ground-state phase diagram is modified by magnetization, with particular emphasis on its impact on the stability and structure of the CFSF and MI phases.
The suppression of CFSF phases at low chemical potentials and the emergence of regions where $a$-SF and $b$-MI coexist highlight the intricate interplay between interspecies interactions and magnetization constraints. Note that this region where the MI and SF phases coexist at once is characteristic to a supersolid phase~\cite{Recati_2023}.

The mean-field approach can be extended to more complex systems incorporating additional interaction terms with additional constants of motion. Similar analyses to those presented in Ref.~\cite{Johnstone2019} can be carried out for models including nearest-neighbor interactions, density-dependent tunneling, or pair tunneling processes, offering a promising direction for future research.

Our results underscore the importance of considering conserved quantities, for example magnetization, and suggest new pathways for engineering complex quantum phases through control of spin or component imbalances.

\section*{ACKNOWLEDGMENTS}
H.D. gratefully acknowledge discussions with P. Rogowski.
This work was supported by the DAINA project of the Polish National Science Center UMO-2020/38/L/ST2/00375.

O.S.\ performed analytical calculations. H.D.\ performed numerical simulation, gather data and prepared all figures. E.W.\ conceived the idea and guided the research. All Authors contributed in the analysis of the results, manuscript preparation and revision.

\bibliography{biblio.bib}

%apsrev4-2.bst 2019-01-14 (MD) hand-edited version of apsrev4-1.bst
%Control: key (0)
%Control: author (8) initials jnrlst
%Control: editor formatted (1) identically to author
%Control: production of article title (0) allowed
%Control: page (0) single
%Control: year (1) truncated
%Control: production of eprint (0) enabled
\begin{thebibliography}{32}%
\makeatletter
\providecommand \@ifxundefined [1]{%
 \@ifx{#1\undefined}
}%
\providecommand \@ifnum [1]{%
 \ifnum #1\expandafter \@firstoftwo
 \else \expandafter \@secondoftwo
 \fi
}%
\providecommand \@ifx [1]{%
 \ifx #1\expandafter \@firstoftwo
 \else \expandafter \@secondoftwo
 \fi
}%
\providecommand \natexlab [1]{#1}%
\providecommand \enquote  [1]{``#1''}%
\providecommand \bibnamefont  [1]{#1}%
\providecommand \bibfnamefont [1]{#1}%
\providecommand \citenamefont [1]{#1}%
\providecommand \href@noop [0]{\@secondoftwo}%
\providecommand \href [0]{\begingroup \@sanitize@url \@href}%
\providecommand \@href[1]{\@@startlink{#1}\@@href}%
\providecommand \@@href[1]{\endgroup#1\@@endlink}%
\providecommand \@sanitize@url [0]{\catcode `\\12\catcode `\$12\catcode
  `\&12\catcode `\#12\catcode `\^12\catcode `\_12\catcode `\%12\relax}%
\providecommand \@@startlink[1]{}%
\providecommand \@@endlink[0]{}%
\providecommand \url  [0]{\begingroup\@sanitize@url \@url }%
\providecommand \@url [1]{\endgroup\@href {#1}{\urlprefix }}%
\providecommand \urlprefix  [0]{URL }%
\providecommand \Eprint [0]{\href }%
\providecommand \doibase [0]{https://doi.org/}%
\providecommand \selectlanguage [0]{\@gobble}%
\providecommand \bibinfo  [0]{\@secondoftwo}%
\providecommand \bibfield  [0]{\@secondoftwo}%
\providecommand \translation [1]{[#1]}%
\providecommand \BibitemOpen [0]{}%
\providecommand \bibitemStop [0]{}%
\providecommand \bibitemNoStop [0]{.\EOS\space}%
\providecommand \EOS [0]{\spacefactor3000\relax}%
\providecommand \BibitemShut  [1]{\csname bibitem#1\endcsname}%
\let\auto@bib@innerbib\@empty
%</preamble>
\bibitem [{\citenamefont {Gersch}\ and\ \citenamefont
  {Knollman}(1963)}]{PhysRev.129.959}%
  \BibitemOpen
  \bibfield  {author} {\bibinfo {author} {\bibfnamefont {H.~A.}\ \bibnamefont
  {Gersch}}\ and\ \bibinfo {author} {\bibfnamefont {G.~C.}\ \bibnamefont
  {Knollman}},\ }\bibfield  {title} {\bibinfo {title} {Quantum cell model for
  bosons},\ }\href {https://doi.org/10.1103/PhysRev.129.959} {\bibfield
  {journal} {\bibinfo  {journal} {Phys. Rev.}\ }\textbf {\bibinfo {volume}
  {129}},\ \bibinfo {pages} {959} (\bibinfo {year} {1963})}\BibitemShut
  {NoStop}%
\bibitem [{\citenamefont {Hubbard}(1963)}]{Hubbard1963}%
  \BibitemOpen
  \bibfield  {author} {\bibinfo {author} {\bibfnamefont {J.}~\bibnamefont
  {Hubbard}},\ }\bibfield  {title} {\bibinfo {title} {Electron correlations in
  narrow energy bands},\ }\href {https://doi.org/10.1098/rspa.1963.0204}
  {\bibfield  {journal} {\bibinfo  {journal} {Proceedings of the Royal Society
  of London. Series A. Mathematical and Physical Sciences}\ }\textbf {\bibinfo
  {volume} {276}},\ \bibinfo {pages} {238} (\bibinfo {year}
  {1963})}\BibitemShut {NoStop}%
\bibitem [{\citenamefont {Hubbard}(1964)}]{Hubbard1964}%
  \BibitemOpen
  \bibfield  {author} {\bibinfo {author} {\bibfnamefont {J.}~\bibnamefont
  {Hubbard}},\ }\bibfield  {title} {\bibinfo {title} {Electron correlations in
  narrow energy bands. ii. the degenerate band case},\ }\href
  {https://doi.org/10.1098/rspa.1964.0019} {\bibfield  {journal} {\bibinfo
  {journal} {Proceedings of the Royal Society of London. Series A. Mathematical
  and Physical Sciences}\ }\textbf {\bibinfo {volume} {277}},\ \bibinfo {pages}
  {237} (\bibinfo {year} {1964})}\BibitemShut {NoStop}%
\bibitem [{\citenamefont {Hubbard}(1965)}]{Hubbard1965}%
  \BibitemOpen
  \bibfield  {author} {\bibinfo {author} {\bibfnamefont {J.}~\bibnamefont
  {Hubbard}},\ }\bibfield  {title} {\bibinfo {title} {Electron correlations in
  narrow energy bands. iii. an improved solution},\ }\href
  {https://doi.org/10.1098/rspa.1964.0190} {\bibfield  {journal} {\bibinfo
  {journal} {Proceedings of the Royal Society of London. Series A. Mathematical
  and Physical Sciences}\ }\textbf {\bibinfo {volume} {281}},\ \bibinfo {pages}
  {401} (\bibinfo {year} {1965})}\BibitemShut {NoStop}%
\bibitem [{\citenamefont {Montorsi}(1992)}]{Montorsi1992}%
  \BibitemOpen
  \bibinfo {editor} {\bibfnamefont {A.}~\bibnamefont {Montorsi}},\ ed.,\
  \href@noop {} {\emph {\bibinfo {title} {The Hubbard Model: A Reprint
  Volume}}}\ (\bibinfo  {publisher} {World Scientific},\ \bibinfo {year}
  {1992})\BibitemShut {NoStop}%
\bibitem [{\citenamefont {Pavarini}\ \emph {et~al.}(2017)\citenamefont
  {Pavarini}, \citenamefont {Koch}, \citenamefont {Scalettar},\ and\
  \citenamefont {Martin}}]{Pavarini2017}%
  \BibitemOpen
  \bibinfo {editor} {\bibfnamefont {E.}~\bibnamefont {Pavarini}}, \bibinfo
  {editor} {\bibfnamefont {E.}~\bibnamefont {Koch}}, \bibinfo {editor}
  {\bibfnamefont {R.}~\bibnamefont {Scalettar}},\ and\ \bibinfo {editor}
  {\bibfnamefont {R.}~\bibnamefont {Martin}},\ eds.,\ \href
  {https://www.cond-mat.de/events/correl17/manuscripts/pavarini.pdf} {\emph
  {\bibinfo {title} {The Physics of Correlated Insulators, Metals, and
  Superconductors}}}\ (\bibinfo  {publisher} {Forschungszentrum Jülich},\
  \bibinfo {year} {2017})\BibitemShut {NoStop}%
\bibitem [{\citenamefont {El-Batanouny}(2020)}]{Majlis2007}%
  \BibitemOpen
  \bibfield  {author} {\bibinfo {author} {\bibfnamefont {M.}~\bibnamefont
  {El-Batanouny}},\ }\href@noop {} {\emph {\bibinfo {title} {Advanced Quantum
  Condensed Matter Physics: One-Body, Many-Body, and Topological
  Perspectives}}}\ (\bibinfo  {publisher} {Cambridge University Press},\
  \bibinfo {year} {2020})\BibitemShut {NoStop}%
\bibitem [{\citenamefont {Pavarini}\ \emph {et~al.}(2015)\citenamefont
  {Pavarini}, \citenamefont {Koch},\ and\ \citenamefont
  {Coleman}}]{Pavarini2015}%
  \BibitemOpen
  \bibinfo {editor} {\bibfnamefont {E.}~\bibnamefont {Pavarini}}, \bibinfo
  {editor} {\bibfnamefont {E.}~\bibnamefont {Koch}},\ and\ \bibinfo {editor}
  {\bibfnamefont {P.}~\bibnamefont {Coleman}},\ eds.,\ \href
  {https://www.cond-mat.de/events/correl15/manuscripts/mielke.pdf} {\emph
  {\bibinfo {title} {Many-Body Physics: From Kondo to Hubbard}}}\ (\bibinfo
  {publisher} {Forschungszentrum Jülich},\ \bibinfo {year} {2015})\BibitemShut
  {NoStop}%
\bibitem [{\citenamefont {Savary}\ and\ \citenamefont
  {Balents}(2017)}]{Savary2017}%
  \BibitemOpen
  \bibfield  {author} {\bibinfo {author} {\bibfnamefont {L.}~\bibnamefont
  {Savary}}\ and\ \bibinfo {author} {\bibfnamefont {L.}~\bibnamefont
  {Balents}},\ }\bibfield  {title} {\bibinfo {title} {Quantum spin liquids: a
  review},\ }\href {https://doi.org/10.1088/0034-4885/80/1/016502} {\bibfield
  {journal} {\bibinfo  {journal} {Reports on Progress in Physics}\ }\textbf
  {\bibinfo {volume} {80}},\ \bibinfo {pages} {016502} (\bibinfo {year}
  {2017})}\BibitemShut {NoStop}%
\bibitem [{\citenamefont {Schollw{\"o}ck}\ \emph {et~al.}(2004)\citenamefont
  {Schollw{\"o}ck}, \citenamefont {Richter}, \citenamefont {Farnell},\ and\
  \citenamefont {Bishop}}]{schollwock2004quantum}%
  \BibitemOpen
  \bibinfo {editor} {\bibfnamefont {U.}~\bibnamefont {Schollw{\"o}ck}},
  \bibinfo {editor} {\bibfnamefont {J.}~\bibnamefont {Richter}}, \bibinfo
  {editor} {\bibfnamefont {D.~J.}\ \bibnamefont {Farnell}},\ and\ \bibinfo
  {editor} {\bibfnamefont {R.~F.}\ \bibnamefont {Bishop}},\ eds.,\ \href
  {https://doi.org/10.1007/b98173} {\emph {\bibinfo {title} {Quantum
  Magnetism}}},\ \bibinfo {series} {Lecture Notes in Physics}, Vol.\ \bibinfo
  {volume} {645}\ (\bibinfo  {publisher} {Springer},\ \bibinfo {year}
  {2004})\BibitemShut {NoStop}%
\bibitem [{\citenamefont {Bloch}\ \emph {et~al.}(2012)\citenamefont {Bloch},
  \citenamefont {Dalibard},\ and\ \citenamefont {Nascimbene}}]{Bloch2012}%
  \BibitemOpen
  \bibfield  {author} {\bibinfo {author} {\bibfnamefont {I.}~\bibnamefont
  {Bloch}}, \bibinfo {author} {\bibfnamefont {J.}~\bibnamefont {Dalibard}},\
  and\ \bibinfo {author} {\bibfnamefont {S.}~\bibnamefont {Nascimbene}},\
  }\bibfield  {title} {\bibinfo {title} {Quantum simulations with ultracold
  quantum gases},\ }\href {https://doi.org/10.1038/nphys2259} {\bibfield
  {journal} {\bibinfo  {journal} {Nat Phys}\ }\textbf {\bibinfo {volume} {8}},\
  \bibinfo {pages} {267} (\bibinfo {year} {2012})}\BibitemShut {NoStop}%
\bibitem [{\citenamefont {Lewenstein}\ \emph {et~al.}(2012)\citenamefont
  {Lewenstein}, \citenamefont {Sanpera},\ and\ \citenamefont
  {Vedral}}]{Lewenstein2012}%
  \BibitemOpen
  \bibfield  {author} {\bibinfo {author} {\bibfnamefont {M.}~\bibnamefont
  {Lewenstein}}, \bibinfo {author} {\bibfnamefont {A.}~\bibnamefont
  {Sanpera}},\ and\ \bibinfo {author} {\bibfnamefont {V.}~\bibnamefont
  {Vedral}},\ }\href@noop {} {\emph {\bibinfo {title} {Ultracold Atoms in
  Optical Lattices: Simulating Quantum Many-Body Systems}}}\ (\bibinfo
  {publisher} {Oxford University Press},\ \bibinfo {address} {Oxford},\
  \bibinfo {year} {2012})\BibitemShut {NoStop}%
\bibitem [{\citenamefont {Dutta}\ \emph {et~al.}(2015)\citenamefont {Dutta},
  \citenamefont {Gajda}, \citenamefont {Hauke}, \citenamefont {Lewenstein},
  \citenamefont {Lühmann}, \citenamefont {Malomed}, \citenamefont
  {Sowiński},\ and\ \citenamefont {Zakrzewski}}]{Dutta_2015}%
  \BibitemOpen
  \bibfield  {author} {\bibinfo {author} {\bibfnamefont {O.}~\bibnamefont
  {Dutta}}, \bibinfo {author} {\bibfnamefont {M.}~\bibnamefont {Gajda}},
  \bibinfo {author} {\bibfnamefont {P.}~\bibnamefont {Hauke}}, \bibinfo
  {author} {\bibfnamefont {M.}~\bibnamefont {Lewenstein}}, \bibinfo {author}
  {\bibfnamefont {D.-S.}\ \bibnamefont {Lühmann}}, \bibinfo {author}
  {\bibfnamefont {B.~A.}\ \bibnamefont {Malomed}}, \bibinfo {author}
  {\bibfnamefont {T.}~\bibnamefont {Sowiński}},\ and\ \bibinfo {author}
  {\bibfnamefont {J.}~\bibnamefont {Zakrzewski}},\ }\bibfield  {title}
  {\bibinfo {title} {Non-standard hubbard models in optical lattices: a
  review},\ }\href {https://doi.org/10.1088/0034-4885/78/6/066001} {\bibfield
  {journal} {\bibinfo  {journal} {Reports on Progress in Physics}\ }\textbf
  {\bibinfo {volume} {78}},\ \bibinfo {pages} {066001} (\bibinfo {year}
  {2015})}\BibitemShut {NoStop}%
\bibitem [{\citenamefont {Kovrizhin}\ \emph {et~al.}(2005)\citenamefont
  {Kovrizhin}, \citenamefont {Pai},\ and\ \citenamefont
  {Sinha}}]{Kovrizhin_2005}%
  \BibitemOpen
  \bibfield  {author} {\bibinfo {author} {\bibfnamefont {D.~L.}\ \bibnamefont
  {Kovrizhin}}, \bibinfo {author} {\bibfnamefont {G.~V.}\ \bibnamefont {Pai}},\
  and\ \bibinfo {author} {\bibfnamefont {S.}~\bibnamefont {Sinha}},\ }\bibfield
   {title} {\bibinfo {title} {Density wave and supersolid phases of correlated
  bosons in an optical lattice},\ }\href
  {https://doi.org/10.1209/epl/i2005-10231-y} {\bibfield  {journal} {\bibinfo
  {journal} {Europhysics Letters}\ }\textbf {\bibinfo {volume} {72}},\ \bibinfo
  {pages} {162} (\bibinfo {year} {2005})}\BibitemShut {NoStop}%
\bibitem [{\citenamefont {Aidelsburger}\ \emph {et~al.}(2013)\citenamefont
  {Aidelsburger}, \citenamefont {Atala}, \citenamefont {Lohse}, \citenamefont
  {Barreiro}, \citenamefont {Paredes},\ and\ \citenamefont
  {Bloch}}]{Aidelsburger2013}%
  \BibitemOpen
  \bibfield  {author} {\bibinfo {author} {\bibfnamefont {M.}~\bibnamefont
  {Aidelsburger}}, \bibinfo {author} {\bibfnamefont {M.}~\bibnamefont {Atala}},
  \bibinfo {author} {\bibfnamefont {M.}~\bibnamefont {Lohse}}, \bibinfo
  {author} {\bibfnamefont {J.~T.}\ \bibnamefont {Barreiro}}, \bibinfo {author}
  {\bibfnamefont {B.}~\bibnamefont {Paredes}},\ and\ \bibinfo {author}
  {\bibfnamefont {I.}~\bibnamefont {Bloch}},\ }\bibfield  {title} {\bibinfo
  {title} {Realization of the hofstadter hamiltonian with ultracold atoms in
  optical lattices},\ }\href {https://doi.org/10.1103/PhysRevLett.111.185301}
  {\bibfield  {journal} {\bibinfo  {journal} {Phys. Rev. Lett.}\ }\textbf
  {\bibinfo {volume} {111}},\ \bibinfo {pages} {185301} (\bibinfo {year}
  {2013})}\BibitemShut {NoStop}%
\bibitem [{\citenamefont {Goldman}\ \emph {et~al.}(2012)\citenamefont
  {Goldman}, \citenamefont {Beugnon},\ and\ \citenamefont
  {Gerbier}}]{PhysRevLett.108.255303}%
  \BibitemOpen
  \bibfield  {author} {\bibinfo {author} {\bibfnamefont {N.}~\bibnamefont
  {Goldman}}, \bibinfo {author} {\bibfnamefont {J.}~\bibnamefont {Beugnon}},\
  and\ \bibinfo {author} {\bibfnamefont {F.}~\bibnamefont {Gerbier}},\
  }\bibfield  {title} {\bibinfo {title} {Detecting chiral edge states in the
  hofstadter optical lattice},\ }\href
  {https://doi.org/10.1103/PhysRevLett.108.255303} {\bibfield  {journal}
  {\bibinfo  {journal} {Phys. Rev. Lett.}\ }\textbf {\bibinfo {volume} {108}},\
  \bibinfo {pages} {255303} (\bibinfo {year} {2012})}\BibitemShut {NoStop}%
\bibitem [{\citenamefont {Bourgund}\ \emph {et~al.}(2025)\citenamefont
  {Bourgund}, \citenamefont {Chalopin}, \citenamefont {Bojovi{\'{c}}},
  \citenamefont {Schl{\"o}mer}, \citenamefont {Wang}, \citenamefont {Franz},
  \citenamefont {Hirthe}, \citenamefont {Bohrdt}, \citenamefont {Grusdt},
  \citenamefont {Bloch},\ and\ \citenamefont {Hilker}}]{Bourgund2025}%
  \BibitemOpen
  \bibfield  {author} {\bibinfo {author} {\bibfnamefont {D.}~\bibnamefont
  {Bourgund}}, \bibinfo {author} {\bibfnamefont {T.}~\bibnamefont {Chalopin}},
  \bibinfo {author} {\bibfnamefont {P.}~\bibnamefont {Bojovi{\'{c}}}}, \bibinfo
  {author} {\bibfnamefont {H.}~\bibnamefont {Schl{\"o}mer}}, \bibinfo {author}
  {\bibfnamefont {S.}~\bibnamefont {Wang}}, \bibinfo {author} {\bibfnamefont
  {T.}~\bibnamefont {Franz}}, \bibinfo {author} {\bibfnamefont
  {S.}~\bibnamefont {Hirthe}}, \bibinfo {author} {\bibfnamefont
  {A.}~\bibnamefont {Bohrdt}}, \bibinfo {author} {\bibfnamefont
  {F.}~\bibnamefont {Grusdt}}, \bibinfo {author} {\bibfnamefont
  {I.}~\bibnamefont {Bloch}},\ and\ \bibinfo {author} {\bibfnamefont {T.~A.}\
  \bibnamefont {Hilker}},\ }\bibfield  {title} {\bibinfo {title} {Formation of
  individual stripes in a mixed-dimensional cold-atom fermi--hubbard system},\
  }\href {https://doi.org/10.1038/s41586-024-08270-7} {\bibfield  {journal}
  {\bibinfo  {journal} {Nature}\ }\textbf {\bibinfo {volume} {637}},\ \bibinfo
  {pages} {57} (\bibinfo {year} {2025})}\BibitemShut {NoStop}%
\bibitem [{\citenamefont {Weckesser}\ \emph {et~al.}(2024)\citenamefont
  {Weckesser}, \citenamefont {Srakaew}, \citenamefont {Blatz}, \citenamefont
  {Wei}, \citenamefont {Adler}, \citenamefont {Agrawal}, \citenamefont
  {Bohrdt}, \citenamefont {Bloch},\ and\ \citenamefont
  {Zeiher}}]{weckesser2024realizationrydbergdressedextendedbose}%
  \BibitemOpen
  \bibfield  {author} {\bibinfo {author} {\bibfnamefont {P.}~\bibnamefont
  {Weckesser}}, \bibinfo {author} {\bibfnamefont {K.}~\bibnamefont {Srakaew}},
  \bibinfo {author} {\bibfnamefont {T.}~\bibnamefont {Blatz}}, \bibinfo
  {author} {\bibfnamefont {D.}~\bibnamefont {Wei}}, \bibinfo {author}
  {\bibfnamefont {D.}~\bibnamefont {Adler}}, \bibinfo {author} {\bibfnamefont
  {S.}~\bibnamefont {Agrawal}}, \bibinfo {author} {\bibfnamefont
  {A.}~\bibnamefont {Bohrdt}}, \bibinfo {author} {\bibfnamefont
  {I.}~\bibnamefont {Bloch}},\ and\ \bibinfo {author} {\bibfnamefont
  {J.}~\bibnamefont {Zeiher}},\ }\href {https://arxiv.org/abs/2405.20128}
  {\bibinfo {title} {Realization of a rydberg-dressed extended bose hubbard
  model}} (\bibinfo {year} {2024}),\ \Eprint {https://arxiv.org/abs/2405.20128}
  {arXiv:2405.20128 [cond-mat.quant-gas]} \BibitemShut {NoStop}%
\bibitem [{\citenamefont {Michel}\ \emph {et~al.}(2024)\citenamefont {Michel},
  \citenamefont {Henriet}, \citenamefont {Domain}, \citenamefont {Browaeys},\
  and\ \citenamefont {Ayral}}]{Browaeys2020}%
  \BibitemOpen
  \bibfield  {author} {\bibinfo {author} {\bibfnamefont {A.}~\bibnamefont
  {Michel}}, \bibinfo {author} {\bibfnamefont {L.}~\bibnamefont {Henriet}},
  \bibinfo {author} {\bibfnamefont {C.}~\bibnamefont {Domain}}, \bibinfo
  {author} {\bibfnamefont {A.}~\bibnamefont {Browaeys}},\ and\ \bibinfo
  {author} {\bibfnamefont {T.}~\bibnamefont {Ayral}},\ }\bibfield  {title}
  {\bibinfo {title} {Hubbard physics with rydberg atoms: Using a quantum spin
  simulator to simulate strong fermionic correlations},\ }\href
  {https://doi.org/10.1103/PhysRevB.109.174409} {\bibfield  {journal} {\bibinfo
   {journal} {Phys. Rev. B}\ }\textbf {\bibinfo {volume} {109}},\ \bibinfo
  {pages} {174409} (\bibinfo {year} {2024})}\BibitemShut {NoStop}%
\bibitem [{\citenamefont {Bernien}\ \emph {et~al.}(2017)\citenamefont
  {Bernien}, \citenamefont {Schwartz}, \citenamefont {Keesling}, \citenamefont
  {Levine}, \citenamefont {Omran}, \citenamefont {Pichler}, \citenamefont
  {Choi}, \citenamefont {Zibrov}, \citenamefont {Endres}, \citenamefont
  {Greiner}, \citenamefont {Vuleti{\'{c}}},\ and\ \citenamefont
  {Lukin}}]{Moses2017}%
  \BibitemOpen
  \bibfield  {author} {\bibinfo {author} {\bibfnamefont {H.}~\bibnamefont
  {Bernien}}, \bibinfo {author} {\bibfnamefont {S.}~\bibnamefont {Schwartz}},
  \bibinfo {author} {\bibfnamefont {A.}~\bibnamefont {Keesling}}, \bibinfo
  {author} {\bibfnamefont {H.}~\bibnamefont {Levine}}, \bibinfo {author}
  {\bibfnamefont {A.}~\bibnamefont {Omran}}, \bibinfo {author} {\bibfnamefont
  {H.}~\bibnamefont {Pichler}}, \bibinfo {author} {\bibfnamefont
  {S.}~\bibnamefont {Choi}}, \bibinfo {author} {\bibfnamefont {A.~S.}\
  \bibnamefont {Zibrov}}, \bibinfo {author} {\bibfnamefont {M.}~\bibnamefont
  {Endres}}, \bibinfo {author} {\bibfnamefont {M.}~\bibnamefont {Greiner}},
  \bibinfo {author} {\bibfnamefont {V.}~\bibnamefont {Vuleti{\'{c}}}},\ and\
  \bibinfo {author} {\bibfnamefont {M.~D.}\ \bibnamefont {Lukin}},\ }\bibfield
  {title} {\bibinfo {title} {Probing many-body dynamics on a 51-atom quantum
  simulator},\ }\href {https://doi.org/10.1038/nature24622} {\bibfield
  {journal} {\bibinfo  {journal} {Nature}\ }\textbf {\bibinfo {volume} {551}},\
  \bibinfo {pages} {579} (\bibinfo {year} {2017})}\BibitemShut {NoStop}%
\bibitem [{\citenamefont {Srinivasan}\ \emph {et~al.}(2024)\citenamefont
  {Srinivasan}, \citenamefont {Beyer}, \citenamefont {Zhu}, \citenamefont
  {Churchill}, \citenamefont {Mehta}, \citenamefont {Sridhar}, \citenamefont
  {Chakrabarti}, \citenamefont {Steuerman}, \citenamefont {Chopra},\ and\
  \citenamefont {Dutt}}]{Britton2012}%
  \BibitemOpen
  \bibfield  {author} {\bibinfo {author} {\bibfnamefont {D.}~\bibnamefont
  {Srinivasan}}, \bibinfo {author} {\bibfnamefont {A.}~\bibnamefont {Beyer}},
  \bibinfo {author} {\bibfnamefont {D.}~\bibnamefont {Zhu}}, \bibinfo {author}
  {\bibfnamefont {S.}~\bibnamefont {Churchill}}, \bibinfo {author}
  {\bibfnamefont {K.}~\bibnamefont {Mehta}}, \bibinfo {author} {\bibfnamefont
  {S.~K.}\ \bibnamefont {Sridhar}}, \bibinfo {author} {\bibfnamefont
  {K.}~\bibnamefont {Chakrabarti}}, \bibinfo {author} {\bibfnamefont {D.~W.}\
  \bibnamefont {Steuerman}}, \bibinfo {author} {\bibfnamefont {N.}~\bibnamefont
  {Chopra}},\ and\ \bibinfo {author} {\bibfnamefont {A.}~\bibnamefont {Dutt}},\
  }\bibfield  {title} {\bibinfo {title} {Trapped-ion quantum simulation of the
  fermi-hubbard model as a lattice gauge theory using hardware-aware native
  gates},\ }\href {https://arxiv.org/abs/2411.07778} {\  (\bibinfo {year}
  {2024})},\ \Eprint {https://arxiv.org/abs/2411.07778} {arXiv:2411.07778
  [quant-ph]} \BibitemShut {NoStop}%
\bibitem [{\citenamefont {Gorshkov}\ \emph {et~al.}(2011)\citenamefont
  {Gorshkov}, \citenamefont {Manmana}, \citenamefont {Chen}, \citenamefont
  {Demler}, \citenamefont {Lukin},\ and\ \citenamefont {Rey}}]{Gorshkov2011}%
  \BibitemOpen
  \bibfield  {author} {\bibinfo {author} {\bibfnamefont {A.~V.}\ \bibnamefont
  {Gorshkov}}, \bibinfo {author} {\bibfnamefont {S.~R.}\ \bibnamefont
  {Manmana}}, \bibinfo {author} {\bibfnamefont {G.}~\bibnamefont {Chen}},
  \bibinfo {author} {\bibfnamefont {E.}~\bibnamefont {Demler}}, \bibinfo
  {author} {\bibfnamefont {M.~D.}\ \bibnamefont {Lukin}},\ and\ \bibinfo
  {author} {\bibfnamefont {A.~M.}\ \bibnamefont {Rey}},\ }\bibfield  {title}
  {\bibinfo {title} {Quantum magnetism with polar alkali-metal dimers},\ }\href
  {https://doi.org/10.1103/PhysRevA.84.033619} {\bibfield  {journal} {\bibinfo
  {journal} {Phys. Rev. A}\ }\textbf {\bibinfo {volume} {84}},\ \bibinfo
  {pages} {033619} (\bibinfo {year} {2011})}\BibitemShut {NoStop}%
\bibitem [{\citenamefont {Tarruell}\ and\ \citenamefont
  {Sanchez-Palencia}(2018)}]{Tarruell2018}%
  \BibitemOpen
  \bibfield  {author} {\bibinfo {author} {\bibfnamefont {L.}~\bibnamefont
  {Tarruell}}\ and\ \bibinfo {author} {\bibfnamefont {L.}~\bibnamefont
  {Sanchez-Palencia}},\ }\bibfield  {title} {\bibinfo {title} {Quantum
  simulation of the hubbard model with ultracold fermions in optical
  lattices},\ }\href
  {https://doi.org/https://doi.org/10.1016/j.crhy.2018.10.013} {\bibfield
  {journal} {\bibinfo  {journal} {Comptes Rendus Physique}\ }\textbf {\bibinfo
  {volume} {19}},\ \bibinfo {pages} {365} (\bibinfo {year} {2018})},\ \bibinfo
  {note} {quantum simulation / Simulation quantique}\BibitemShut {NoStop}%
\bibitem [{\citenamefont {Altman}\ \emph {et~al.}(2003)\citenamefont {Altman},
  \citenamefont {Hofstetter}, \citenamefont {Demler},\ and\ \citenamefont
  {Lukin}}]{Altman_2003}%
  \BibitemOpen
  \bibfield  {author} {\bibinfo {author} {\bibfnamefont {E.}~\bibnamefont
  {Altman}}, \bibinfo {author} {\bibfnamefont {W.}~\bibnamefont {Hofstetter}},
  \bibinfo {author} {\bibfnamefont {E.}~\bibnamefont {Demler}},\ and\ \bibinfo
  {author} {\bibfnamefont {M.~D.}\ \bibnamefont {Lukin}},\ }\bibfield  {title}
  {\bibinfo {title} {Phase diagram of two-component bosons on an optical
  lattice},\ }\href {https://doi.org/10.1088/1367-2630/5/1/113} {\bibfield
  {journal} {\bibinfo  {journal} {New Journal of Physics}\ }\textbf {\bibinfo
  {volume} {5}},\ \bibinfo {pages} {113} (\bibinfo {year} {2003})}\BibitemShut
  {NoStop}%
\bibitem [{\citenamefont {Kuklov}\ \emph {et~al.}(2004)\citenamefont {Kuklov},
  \citenamefont {Prokof'ev},\ and\ \citenamefont
  {Svistunov}}]{PhysRevLett.92.050402}%
  \BibitemOpen
  \bibfield  {author} {\bibinfo {author} {\bibfnamefont {A.}~\bibnamefont
  {Kuklov}}, \bibinfo {author} {\bibfnamefont {N.}~\bibnamefont {Prokof'ev}},\
  and\ \bibinfo {author} {\bibfnamefont {B.}~\bibnamefont {Svistunov}},\
  }\bibfield  {title} {\bibinfo {title} {Commensurate two-component bosons in
  an optical lattice: Ground state phase diagram},\ }\href
  {https://doi.org/10.1103/PhysRevLett.92.050402} {\bibfield  {journal}
  {\bibinfo  {journal} {Phys. Rev. Lett.}\ }\textbf {\bibinfo {volume} {92}},\
  \bibinfo {pages} {050402} (\bibinfo {year} {2004})}\BibitemShut {NoStop}%
\bibitem [{\citenamefont {Kuklov}\ and\ \citenamefont
  {Svistunov}(2003)}]{PhysRevLett.90.100401}%
  \BibitemOpen
  \bibfield  {author} {\bibinfo {author} {\bibfnamefont {A.~B.}\ \bibnamefont
  {Kuklov}}\ and\ \bibinfo {author} {\bibfnamefont {B.~V.}\ \bibnamefont
  {Svistunov}},\ }\bibfield  {title} {\bibinfo {title} {Counterflow
  superfluidity of two-species ultracold atoms in a commensurate optical
  lattice},\ }\href {https://doi.org/10.1103/PhysRevLett.90.100401} {\bibfield
  {journal} {\bibinfo  {journal} {Phys. Rev. Lett.}\ }\textbf {\bibinfo
  {volume} {90}},\ \bibinfo {pages} {100401} (\bibinfo {year}
  {2003})}\BibitemShut {NoStop}%
\bibitem [{\citenamefont {Hu}\ \emph {et~al.}(2009)\citenamefont {Hu},
  \citenamefont {Mathey}, \citenamefont {Danshita}, \citenamefont {Tiesinga},
  \citenamefont {Williams},\ and\ \citenamefont {Clark}}]{PhysRevA.80.023619}%
  \BibitemOpen
  \bibfield  {author} {\bibinfo {author} {\bibfnamefont {A.}~\bibnamefont
  {Hu}}, \bibinfo {author} {\bibfnamefont {L.}~\bibnamefont {Mathey}}, \bibinfo
  {author} {\bibfnamefont {I.}~\bibnamefont {Danshita}}, \bibinfo {author}
  {\bibfnamefont {E.}~\bibnamefont {Tiesinga}}, \bibinfo {author}
  {\bibfnamefont {C.~J.}\ \bibnamefont {Williams}},\ and\ \bibinfo {author}
  {\bibfnamefont {C.~W.}\ \bibnamefont {Clark}},\ }\bibfield  {title} {\bibinfo
  {title} {Counterflow and paired superfluidity in one-dimensional bose
  mixtures in optical lattices},\ }\href
  {https://doi.org/10.1103/PhysRevA.80.023619} {\bibfield  {journal} {\bibinfo
  {journal} {Phys. Rev. A}\ }\textbf {\bibinfo {volume} {80}},\ \bibinfo
  {pages} {023619} (\bibinfo {year} {2009})}\BibitemShut {NoStop}%
\bibitem [{\citenamefont {Colussi}\ \emph {et~al.}(2022)\citenamefont
  {Colussi}, \citenamefont {Caleffi}, \citenamefont {Menotti},\ and\
  \citenamefont {Recati}}]{10.21468/SciPostPhys.12.3.111}%
  \BibitemOpen
  \bibfield  {author} {\bibinfo {author} {\bibfnamefont {V.~E.}\ \bibnamefont
  {Colussi}}, \bibinfo {author} {\bibfnamefont {F.}~\bibnamefont {Caleffi}},
  \bibinfo {author} {\bibfnamefont {C.}~\bibnamefont {Menotti}},\ and\ \bibinfo
  {author} {\bibfnamefont {A.}~\bibnamefont {Recati}},\ }\bibfield  {title}
  {\bibinfo {title} {{Quantum Gutzwiller approach for the two-component
  Bose-Hubbard model}},\ }\href {https://doi.org/10.21468/SciPostPhys.12.3.111}
  {\bibfield  {journal} {\bibinfo  {journal} {SciPost Phys.}\ }\textbf
  {\bibinfo {volume} {12}},\ \bibinfo {pages} {111} (\bibinfo {year}
  {2022})}\BibitemShut {NoStop}%
\bibitem [{\citenamefont {de~Hond}\ \emph {et~al.}(2022)\citenamefont
  {de~Hond}, \citenamefont {Xiang}, \citenamefont {Chung}, \citenamefont
  {Cruz-Col\'on}, \citenamefont {Chen}, \citenamefont {Burton}, \citenamefont
  {Kennedy},\ and\ \citenamefont {Ketterle}}]{PhysRevLett.128.093401}%
  \BibitemOpen
  \bibfield  {author} {\bibinfo {author} {\bibfnamefont {J.}~\bibnamefont
  {de~Hond}}, \bibinfo {author} {\bibfnamefont {J.}~\bibnamefont {Xiang}},
  \bibinfo {author} {\bibfnamefont {W.~C.}\ \bibnamefont {Chung}}, \bibinfo
  {author} {\bibfnamefont {E.}~\bibnamefont {Cruz-Col\'on}}, \bibinfo {author}
  {\bibfnamefont {W.}~\bibnamefont {Chen}}, \bibinfo {author} {\bibfnamefont
  {W.~C.}\ \bibnamefont {Burton}}, \bibinfo {author} {\bibfnamefont {C.~J.}\
  \bibnamefont {Kennedy}},\ and\ \bibinfo {author} {\bibfnamefont
  {W.}~\bibnamefont {Ketterle}},\ }\bibfield  {title} {\bibinfo {title}
  {Preparation of the spin-mott state: A spinful mott insulator of repulsively
  bound pairs},\ }\href {https://doi.org/10.1103/PhysRevLett.128.093401}
  {\bibfield  {journal} {\bibinfo  {journal} {Phys. Rev. Lett.}\ }\textbf
  {\bibinfo {volume} {128}},\ \bibinfo {pages} {093401} (\bibinfo {year}
  {2022})}\BibitemShut {NoStop}%
\bibitem [{\citenamefont {Jaksch}\ \emph {et~al.}(1998)\citenamefont {Jaksch},
  \citenamefont {Bruder}, \citenamefont {Cirac}, \citenamefont {Gardiner},\
  and\ \citenamefont {Zoller}}]{Jaksch1998-yn}%
  \BibitemOpen
  \bibfield  {author} {\bibinfo {author} {\bibfnamefont {D.}~\bibnamefont
  {Jaksch}}, \bibinfo {author} {\bibfnamefont {C.}~\bibnamefont {Bruder}},
  \bibinfo {author} {\bibfnamefont {J.~I.}\ \bibnamefont {Cirac}}, \bibinfo
  {author} {\bibfnamefont {C.~W.}\ \bibnamefont {Gardiner}},\ and\ \bibinfo
  {author} {\bibfnamefont {P.}~\bibnamefont {Zoller}},\ }\bibfield  {title}
  {\bibinfo {title} {Cold bosonic atoms in optical lattices},\ }\href@noop {}
  {\  (\bibinfo {year} {1998})}\BibitemShut {NoStop}%
\bibitem [{\citenamefont {Recati}\ and\ \citenamefont
  {Stringari}(2023)}]{Recati_2023}%
  \BibitemOpen
  \bibfield  {author} {\bibinfo {author} {\bibfnamefont {A.}~\bibnamefont
  {Recati}}\ and\ \bibinfo {author} {\bibfnamefont {S.}~\bibnamefont
  {Stringari}},\ }\bibfield  {title} {\bibinfo {title} {Supersolidity in
  ultracold dipolar gases},\ }\href
  {https://doi.org/10.1038/s42254-023-00648-2} {\bibfield  {journal} {\bibinfo
  {journal} {Nature Reviews Physics}\ }\textbf {\bibinfo {volume} {5}},\
  \bibinfo {pages} {735–743} (\bibinfo {year} {2023})}\BibitemShut {NoStop}%
\bibitem [{\citenamefont {Johnstone}\ \emph {et~al.}(2019)\citenamefont
  {Johnstone}, \citenamefont {Westerberg}, \citenamefont {Duncan},\ and\
  \citenamefont {\"Ohberg}}]{Johnstone2019}%
  \BibitemOpen
  \bibfield  {author} {\bibinfo {author} {\bibfnamefont {D.}~\bibnamefont
  {Johnstone}}, \bibinfo {author} {\bibfnamefont {N.}~\bibnamefont
  {Westerberg}}, \bibinfo {author} {\bibfnamefont {C.~W.}\ \bibnamefont
  {Duncan}},\ and\ \bibinfo {author} {\bibfnamefont {P.}~\bibnamefont
  {\"Ohberg}},\ }\bibfield  {title} {\bibinfo {title} {Staggered ground states
  in an optical lattice},\ }\href@noop {} {\bibfield  {journal} {\bibinfo
  {journal} {Phys. Rev. A}\ } (\bibinfo {year} {2019})}\BibitemShut {NoStop}%
\end{thebibliography}%
\end{document}